\documentclass[a4paper,12pt]{article}

\usepackage{geometry}
\usepackage{rotating}
\usepackage{color}
\usepackage{comment}
\usepackage{graphicx}
\usepackage{setspace}

\usepackage{epstopdf}
\usepackage{subfig}

\usepackage{amsmath}
\usepackage{amsfonts}
\usepackage{amssymb}

\title{Prediction and replication from case-control sequencing studies using custom genotyping and additional sequencing}
\author{C Ryan King, Paul J Rathouz, Dan L Nicolae}

\makeatletter
\renewcommand{\@biblabel}[1]{\quad#1.}
\makeatother
\newcommand{\E}{\mbox{\rm E}}
\newcommand{\var}{\mbox{\rm var}}

\newcommand{\pr}{\mbox{\rm Pr}}
\newcommand{\logit}{\mbox{\rm logit}}

\begin{document}

\maketitle
\abstract{
We present two results about using allele-count (AC) burdens of rare SNPs discovered in a case-control sequencing study for prediction or validation in an external prospective study.
When genotyping only the SNPs polymorphic in the sequence data, the phenotype to AC correlation tends to be larger in the replication data than the primary study. 
Conversely, if the replication sample is sequenced, ACs of SNPs which are novel in the replication tend to have much smaller or opposite signed associations.
We explain this by first deriving the AC-phenotype association implied by a model of diverse SNP effects, and second accounting for the shifted distribution of SNP effects when using a case-control study as a filter for SNP inclusion.
In rare diseases, the case population is depleted of protective SNPs and enriched for deleterious SNPs, creating the above difference in AC associations. 
This phenomenon is most relevant in re-sequencing for risk prediction in rare diseases with heterogeneous rare mutations because it applies to SNPs with MAF near 1 out of the case-control sample size and is exaggerated when SNP log-odds ratios come from a heavy-tailed distribution.
It also suggests a ``winner's curse'' in which most risk increasing SNPs at a particular MAF are quickly discovered and future sequencing finds more protective or irrelevant SNPs.
%
}

\section{Introduction}
The increasing affordability of genome sequencing has seen caused many investigators to look for rare variation underlying human disease.
In response, gene-level association tests for combining information from rare SNPs have become increasing well understood and widely used.
However, the replication of these associations and the role of rare SNPs discovered in sequencing projects for estimation of risk has been less well defined in the literature.

This paper considers the following hypothetical two-phase research program.
In  phase one, a case-control study without substantial external data is conducted using genomic sequencing over the gene (or region) of interest.
In phase two, a prospective replication sample is drawn from the same parent population without regard to phenotype.
In phase two, genotypes are either assayed on the set of SNPs already found to be polymorphic in phase one (custom genotyping) or assayed on all polymorphic SNPs in the gene (additional sequencing).
We consider a very simple analytical approach to this experiment and derive its properties; of course the analysis we focus on will not be the most relevant for all possible investigators, but we argue below that it is well suited to several goals.
After phase one, the investigator estimates an overall association between rare SNPs and phenotype using logistic regression of a per-person sum of rare alleles versus phenotype, similar to an allelic burden test \cite{madsen_groupwise_2009}.
After the phase two, we consider several options open to the investigator. 
First, she can predict phenotypes using allele counts of the new participants and the associations implied by the phase one data.
Second, if the phenotypes of the new participants are known she can re-estimate the allele-count association and quantitatively compare it to that in the old data in an attempt to validate the association.
Third, if additional sequencing is performed she can estimate an allele-count association with the newly discovered (novel) SNPs and compare it to that of the previously ascertained SNPs (those polymorphic in the phase one data).

Why would an investigator do this?
We argue that this analysis is relevant for looking at very rare SNPs, especially those whose minor allele appears only a few times in the data.
Sequencing studies will continue to discover new rare SNPs for the foreseeable future, and individuals sequenced for risk prediction will routinely carry previously unseen and seldomly observed SNPs \cite{coventry_deep_2010, keinan_recent_2012}; detection of these rare SNPs is the unique advantage of sequencing.
While allele-count tests and their weighted relatives are certainly not the most powerful tests in many circumstances, they do capture most of the information contained in SNPs which appear only once in the data.
Additionally, tests and estimates of the mean SNP effect can be part of more complex procedures \cite{king_evolutionary_2010}.
Predicted phenotypes and allelic burdens offer a transparent and intuitive approach to validation of a prior finding.
From a ``personalized medicine'' prediction perspective, rare SNPs are a substantial contrast to GWAS with large samples and common SNPs; we will never be in the comfortable position of having precise plug-in estimates for each SNP.
The best we can hope for with novel and seldomly observed SNPs is a pooled effect estimate which extrapolates to new samples based on other SNPs in prior data in the same gene.
In particular, prediction from novel SNPs uses \emph{only} information obtained by borrowing from other SNPs.
We will show that the optimal effect estimates for prediction using previously polymorphic and novel SNPs are quite different.
Finally, associating novel SNPs by sequencing the phase two cohort offers a second form of replication that the gene or subunit under study is important.

A crucial feature of our approach is that we think of the log odds-ratios (lORs) of individual SNPs in a gene as coming from an incompletely observed group rather than fixed unknowns which we want to infer.
We also assume that lORs are heterogeneous and described by a continuous distribution, but do not assume prior information on which SNPs are likely to have large effect sizes.
That is, our analysis treats them like ``random effects'' rather than ``fixed effects.''
Our results are obtained by averaging over both the generation of SNP lORs from an underlying distribution and sampling of individuals from a population with SNP effects held fixed.

Based on the classical result that lORs are consistently estimated by both case-control and prospective study designs \cite{prentice_logistic_1979, staicu_equivalence_2010}, one might anticipate that the disease-risk to allele-count lOR for phase two genotyping, re-using the SNPs from the first phase case-control study would coincide on average with parameter estimates from phase one. 
However, we will show that the effect estimates from the two study designs are different in expectation.
One might also expect that the lOR for the count of novel alleles in phase two sequencing would be the same on average as that from phase one, or at least the same direction.
However, we show that the lOR for novel SNPs is often much less or opposite signed.


\section{Notation and Generative Models}\label{notationsec}

\begin{table}[ht]
\begin{center}
\begin{tabular}{|l|l|}
  \hline
N	&number of cases = number of controls\\
M	&number of SNPs in gene \\
i	&index for participants\\
j	&index for SNP\\
$Y_i$	&participant outcome, 1= affected, 0=unaffected	\\
$G_{ij}$	&numeric genotype of person i at SNP j\\
$g_i$	&count of rare alleles for person i,  $g_i=\sum_j G_{ij}$\\
$f_j$	&MAF of SNP j	\\
$K$	&prevalence of disease	\\
$\gamma_j$	&lOR of $j^{th}$ SNP\\
$\mu$	&E$[\gamma]$ considering all SNPs in the gene\\
$\tau^2$	&Var$[\gamma]$ considering all SNPs in the gene\\
$\mu_{p}$	&E$[\gamma | \textrm{SNP polymorphic in primary study}]$\\
$\mu_{r}$	&E$[\gamma | \textrm{SNP novel to replication study}]$\\
$\tau^2_{p}$	&Var$[\gamma | \textrm{SNP polymorphic in primary study}]$\\
$\tau^2_{r}$	&Var$[\gamma | \textrm{SNP novel to replication study}]$\\
$\beta^p_a$	&parameter of model \eqref{PAM}, E[lOR Y vs g in primary study]\\
$\beta^r_a$	&E[lOR] in replication using all SNPs (resequencing)\\
$\beta^r_p$	&E[lOR] in replication, only SNPs polymorphic in primary study\\
$\beta^r_r$	&E[lOR] in replication, only SNPs novel to replication study\\
   \hline
\end{tabular}
\caption{Notation}
\label{NotationTable}
\end{center}
\end{table}

\subsection{Generative and Descriptive Model}
Denote the status of the $i^{th}$ individual in a study by $Y_i =1$ for diseased and $0$ for undiseased. 
In the phase one case-control study, assume an equal number $N$ of diseased cases and disease-free controls.  Let $G$ be a $2N \times M$ matrix of participant genotypes, where $G_{ij} \in \{0,1\}$ for the $i^{th}$ participant having either zero or one minor allele at the $j^{th}$ locus, and $G_i$ the row vector for the $i^{th}$ participant. 
We ignore minor allele homozygotes as they are uncommon at rare SNPs and complicate some of the expressions without meaningfully changing the qualitative or the numerical results.  
To isolate the behavior of SNPs at different minor allele frequencies (MAF), in all models, we fix all SNPs to come from a common MAF $f$, which we will vary between scenarios.

We assume a generative model in which disease status arises via a vector of SNP-specific effects $\gamma_j$. Specifically, 
\begin{equation}\label{ss-m1}
 \logit \left(\pr \lbrace Y_i =1 | G_i , \gamma \rbrace \right) = G_i \gamma + \alpha \ ,
\end{equation}
 where $\gamma_j$ denotes the log-odds ratio of the $j^{th}$ SNP, $\logit()$ is the logistic link function, and $\alpha$ the intercept.
In the population, the entire ensemble of SNP lORs are assumed to arise as independent and identically distributed (IID) effects from a distribution with mean $\E(\gamma) = \mu$ and variance $\var(\gamma) = \tau^2$; we will consider both Gaussian and non-Gaussian distributions.

In phase one, the $2N$ individuals are sampled with equal probability from the affected and unaffected.
In phase two, we envision randomly drawing an individual from the same larger population from which the initial case-control study was sampled; this individual is equipped with a random draw $G_i$ from the population of genotype vectors. 
We keep the vector $\gamma$ of SNP effects (lORs) the same between the case-control and subsequent experiment; however, we do not carry over per-SNP inferences about lORs ($\hat{\gamma}$) from the case-control study to the prediction problem.
The total population is assumed large enough that re-sampling individuals is ignorable.
We will show some results at different phase one sample sizes, but we do not require it to be ``small'' in any absolute way.
We assume that phase one and two sample sizes are large enough that the usual regression asymptotic results apply.
Our results in phase two are averaged over sampling new participants, and can be thought of as close to what one would obtain with ``large'' samples.

We also define the allele-count burden model:
\begin{equation}\label{PAM}
 logit \left(\pr \lbrace Y_i =1 | g_i, \beta \rbrace \right)=  g_i \beta + a , 
\end{equation}
where $g_i$ is a count of minor alleles.
The burden model is descriptive, not generative for the data, and some care is required in interpreting it.
We intend to study ``what one would estimate,'' but the estimated $\hat{\beta}$ in a particular sample is both stochastic from sampling of participants and a function of the realized SNP lORs, the number of SNPs in the population, their MAF, and the sample size of the case-control study (as we will show).
We will show some results in the Supplement for particular realizations, but to avoid dependence on sampling variation we will refer to the ``true'' $\beta$ as the expected value $E_{\gamma}\left[ E_{\textrm{sampling Y, G}}\left[\hat{\beta}|Y,G,\gamma \right] | \mu, \tau\right]$ where the expectation is over both generation of $\gamma$ from its distribution and selection of participants from the larger population.

We operationalize the ``rare SNP effect'' as the lOR $\beta_a^p$ obtained by regressing the disease status versus the number of rare alleles in a \underline{p}rimary case-control sequencing study using \underline{a}ll available SNPs (superscript for study design, subscript for SNP type), and contrast it to the effect observed in future cohort \underline{r}eplication or prediction samples using all SNPs $\beta^r_a$, only SNPs \underline{p}olymorphic in the primary study $\beta^r_p$, and only SNPs novel to the \underline{r}eplication study $\beta^r_r$.
We adopt the same subscript conventions for the moments of the distribution of $\gamma$; $\mu$ and $\tau$ are always the mean and standard deviation of SNP effects in the entire population; the mean of lORs for those SNPs sampled in a primary case-control study is $\mu_{p}$, and those novel in a replication $\mu_{r}$.

We refer to a test of ${\beta}=0$ as an \textit{allele-count test}; for example a t-test of the number of minor alleles between cases and controls.

\section{Results}

\subsection{Allele-count effects are generally non-zero} 
\label{prim-allele-count-effects}
We begin by finding the regression association $\beta$ in equation \eqref{PAM} implied by our random effect model for $\gamma$ in a prospective cohort study design when using all SNPs.

First, re-write the logistic model \eqref{ss-m1} as the equivalent latent liability model,
\begin{equation} \label{logisticnormal}
\pr \lbrace Y_i = 1 | G_i, \mu, \tau \rbrace  = E_\gamma E_{\epsilon_i}\left[ I\left( \epsilon_i + G_i \gamma > c | \gamma \right) \right] 
\end{equation}
where $\epsilon_i$ is a latent variable following a logistic distribution, $I()$ the indicator function, $c=-log(\frac{K}{1-K}) \approx -log(K)$ is the threshold value for the prevalence \textit{K} of the disease among those with no rare alleles, and subscripts on $E$ denote random variables over which expectation is computed.
Next, approximate the logistic variable by a Gaussian scaled by 1.6 \cite{logistic_scale}. 
Utilizing the IID Gaussian assumption of $\gamma_j$, a scalar sum $g_i$ is sufficient for vector $G_i$, yielding
\begin{equation*}
\pr \lbrace Y_i = 1 | G_i, \mu, \tau  \rbrace  \approx  Pr_{Z_i} \lbrace  Z_i \sqrt{1.6^2 + \tau^2 g_i} + \mu g_i > c | g_i\rbrace,
\end{equation*}
where $Z_i$ is a standard normal. Re-apply the Normal-logistic approximation to obtain
\begin{equation*}
\pr \lbrace Y_i = 1 | G_i, \mu, \tau  \rbrace  \approx  Pr_{\epsilon^*_i} \left\{ \epsilon_i^*  + \frac{\mu g_i - c}{\sqrt{1 + \tau^2 g_i /1.6^2}} > 0 \right\},
\end{equation*}
where $\epsilon_i^*$ is again logistic distributed. Finally, returning to a logistic regression form,
\begin{equation}
\label{nonlinear-marginalmodel}
\logit\left( \pr \lbrace Y_i =1 | G_i, \mu, \tau \rbrace \right) \approx
\logit\left( \pr \lbrace Y_i =1 | g_i, \mu, \tau \rbrace \right) \approx \frac{-c + \mu g_{i}}{\sqrt{1 + \tau^2 g_{i}/1.6^2}}\ 
\end{equation}

Expression \eqref{nonlinear-marginalmodel} is non-linear in $g_i$; therefore, in any logistic-linear approximation, $\beta$ will depend on the distribution of $g_i$. 
Although there is no ``typical'' distribution of $g_i$, we focus on cases where for most individuals it is small, and therefore present calculations where $\beta$ reflects the contrast/difference between $g_i=1$ and $g_i=0$.
In that context, approximation \eqref{nonlinear-marginalmodel} can be manipulated to obtain,
\begin{equation}\label{marginalmodel}
\beta \approx \frac{ \mu + c (\sqrt{1 + \tau^2 /1.6^2}-1)} {\sqrt{1 + \tau^2/1.6^2}} \ .
\end{equation}
Contrasts at alternative values of $g_i$ can be obtained trivially.

The quality of approximation \eqref{marginalmodel} depends on disease prevalence because of the tails of the logistic and Gaussian distributions differ; it is acceptable for prevalence ($K$) greater than 5\%, (see Figure \ref{normal_approx_bad}).
The $\beta$ in approximation \eqref{marginalmodel} is a function of the moments of SNP lORs included into $G$ and $g$; as long as the distribution of $\gamma$ approximates a Gaussian it applies equally well to any of the $\beta$ parameters in Table \ref{NotationTable}.

Because we are interested only in rare alleles, in most of our simulation settings (described in the Supplement) most SNPs are sampled a small number of times and $g_{i}$ is also small (often 0 or 1).
As a result individuals are nearly independent of one another, and the mis-specification created in equation \eqref{PAM} by not considering the covariance of $Y$ is minor, and 
the only practically detectable effect of the gene will be the shift in the mean trait value, in contrast to many current sequencing analysis methods which estimate or test the covariance of individuals created by SNP effects \cite{neale_testing_2011, wu_rare-variant_2011, liu_novel_2010, king_evolutionary_2010}.

Expression \eqref{marginalmodel} is linear in the log-prevalence and mean of SNP lORs and non-linear in the variance of SNP lORs.
In generative model~\eqref{ss-m1}, the absence of allelic effects, i.e., $\gamma_j=0$ for all $j$, will clearly yield no association between either $G_i$ or $g_i$ and $Y_i$. We contrast this \emph{strong} null hypothesis of no association with the \emph{weaker} hypothesis wherein the mean effect $\mu=0$, but the variance $\tau^2$ of effects is positive. In this latter case, the non-linearity of~\eqref{nonlinear-marginalmodel} in $g_i$ gives rise to non-zero effects of allele counts.

\begin{figure}[!ht]
\begin{center}
\centerline{\includegraphics[width=6in]{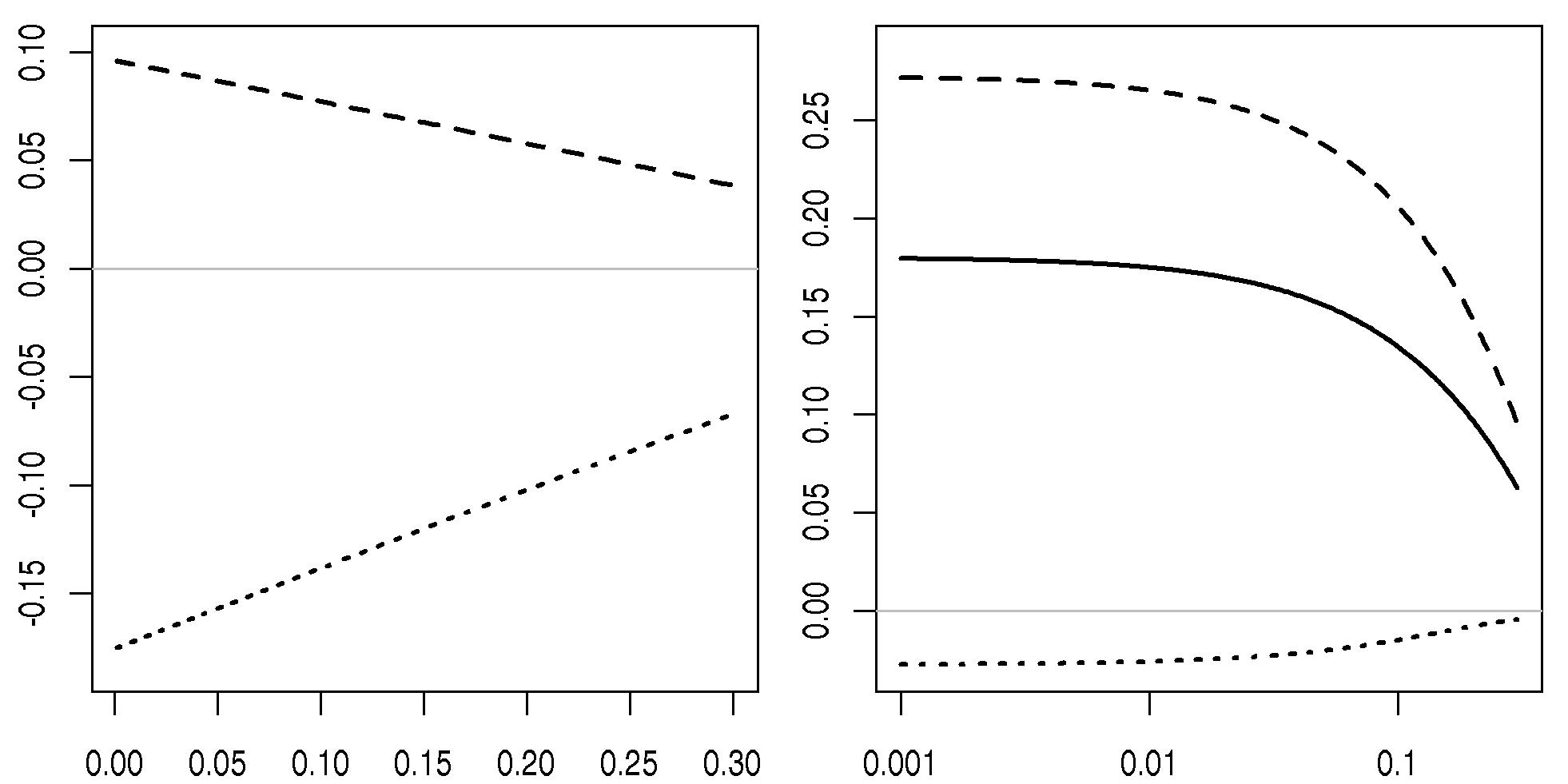}}
\end{center}
\caption{
{\bf Expected lOR and $\beta$ in prospective and case-control studies versus baseline disease prevalence.} Left panel (discussed in Section \ref{ascertainment_bias_section}): mean of previously polymorphic SNP lORs $\mu_{p}$ (dashed) and novel SNP lORs $\mu_{r}$ (dotted) versus disease prevalence.  Right panel: expected burden lORs $\beta_{a}^{r} = \beta_a^{p}$ (solid) versus prevalence (logarithmic x-axis), and from Section \ref{ascertainment_bias_section} $\beta_{r}^{r}$ (dotted), $\beta_{p}^{r}$ (dashed). $\tau=0.6$, $\mu=0$, N=100, $f=.005$. Prevalence defined  among those with no rare minor alleles. Curves obtained by numerical integration of the distribution of SNP effects given polymorphic status using equations \eqref{exposure_pdds}-\eqref{exposure_pdds_ca} and numerical integration of marginal effect using equation \eqref{logisticnormal}.
}
\label{prevelence_curve}
\end{figure}

\begin{figure}[!ht]
\begin{center}
\centerline{\includegraphics[width=6in]{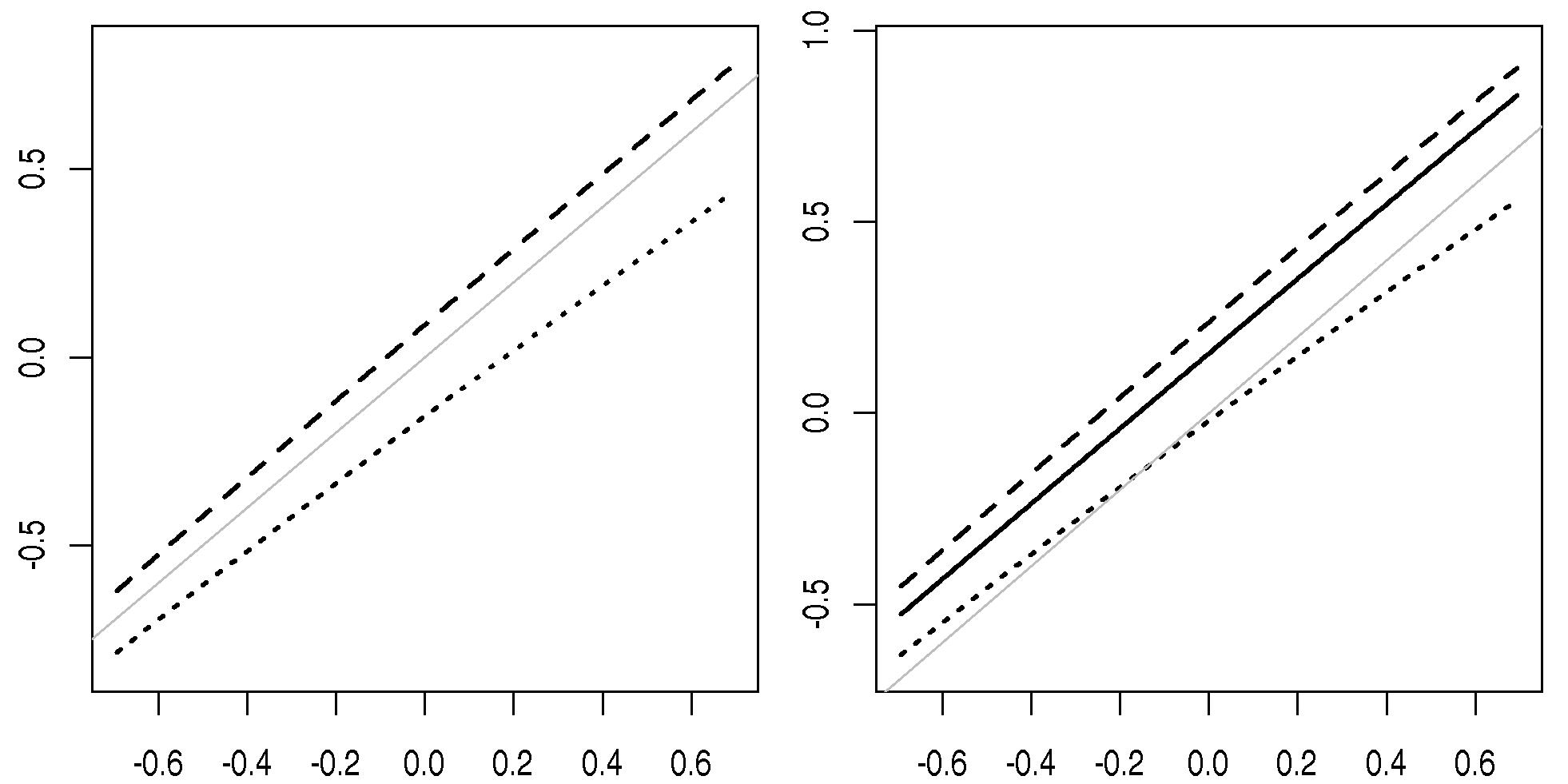}}
\end{center}
\caption{
{\bf Expected lOR and $\beta$ in prospective and case-control studies versus mean lOR.} Left panel: $\mu_{r}$ (dotted) and $\mu_{p}$ (dashed) versus $\mu$.  Right panel: $\beta_{r}^{r}$ (dotted), $\beta_{p}^{r}$ (dashed) and $\beta_{a}^{r} = \beta_a^{p}$ (solid) versus $\mu$. Prevalence 5\%, $\tau=0.6$, N=100, $f=.005$. Curves obtained by numerical integration of the distribution of SNP effects given polymorphic status using equations \eqref{exposure_pdds}-\eqref{exposure_pdds_ca} and numerical integration of marginal effect using equation \eqref{logisticnormal}. Faint gray line of identity
}
\label{mu_curve}
\end{figure}

\begin{figure}[!ht]
\begin{center}
\centerline{\includegraphics[width=6in]{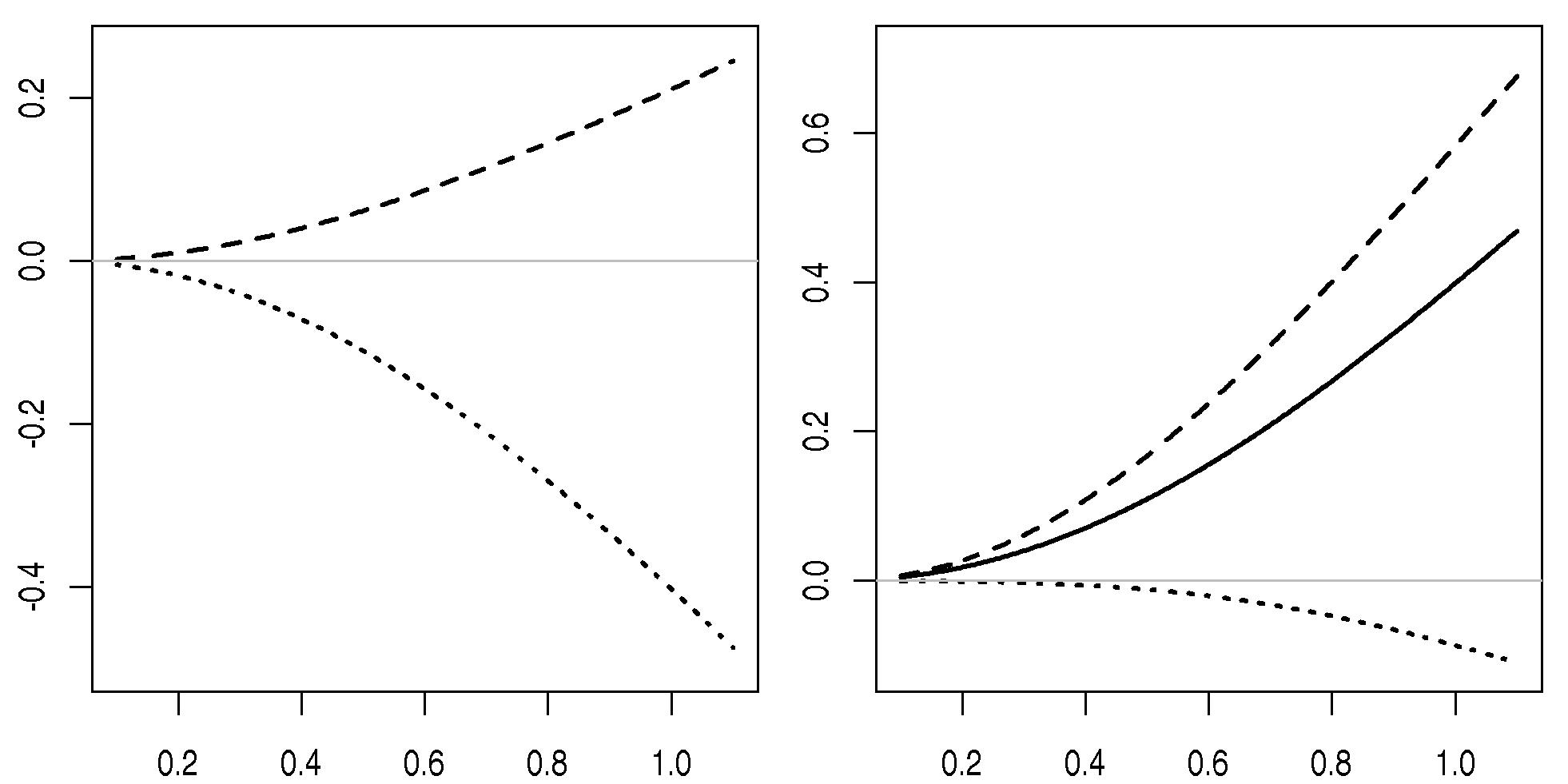}}
\end{center}
\caption{
{\bf Expected lOR and $\beta$ in prospective and case-control studies versus SD of lORs.} Left panel: $\mu_{r}$ (dotted) and $\mu_{p}$ (dashed) versus $\tau$.  Right panel: $\beta_{r}^{r}$ (dotted), $\beta_{p}^{r}$ (dashed) and $\beta_{a}^{r} = \beta_a^{p}$ (solid) versus $\tau$. Prevalence 5\%, $\mu=0$, N=100, $f=.005$. Curves obtained by numerical integration of the distribution of SNP effects given polymorphic status using equations \eqref{exposure_pdds}-\eqref{exposure_pdds_ca} and numerical integration of marginal effect using equation \eqref{logisticnormal}.
}
\label{tau_curve}
\end{figure}

To better illustrate these findings, we present numerical calculations in Figures \ref{prevelence_curve}, \ref{mu_curve}, \ref{tau_curve}, which jointly display results from this section and the next.
All three figures are obtained by numerical integration of equation \eqref{logisticnormal} rather than approximation \eqref{marginalmodel} and show $\beta$ parameters contrasting $g_i=0$ vs $g_i=1$.
The left panels and dashed or dotted curves are obtained using results from Section \ref{ascertainment_bias_section} and discussed there.
The solid curve of the right panel of Figures \ref{prevelence_curve}, \ref{mu_curve}, \ref{tau_curve} display the $\beta_a^{p}$ implied by equation \eqref{logisticnormal} varying (respectively) the prevalence, mean, and standard deviation of SNP lORs.

Figure \ref{prevelence_curve} right panel shows that with symmetric but fairly variable SNP lORs (mean 0, SD 0.6) and uncommon to rare diseases (prevalence less than 5\%) SNP allele counts predict disease status with a non-trivial but small effect sizes (OR 1.15 - 1.20 per allele).
At higher prevalences the allele count lOR decreases proportionally with the negative log of prevalence; at lower values it reaches a plateau.
As mentioned above, the discordance between Figure \ref{prevelence_curve} and approximation \eqref{marginalmodel} (the plateau) is due to the failure of a Gaussian approximation of the tails of a logistic distribution.
Figure \ref{mu_curve} shows that, as one would anticipate, a non-zero average of SNP lORs is a main driver of the expected burden lOR; the two are linearly related with a coefficient just slightly off unity and an offset described in Figure \ref{prevelence_curve}.
Figure \ref{tau_curve} shows the SD of SNP lORs to be another major determinant of $\beta_a^{p}$.
Notably, $\beta_a^{p}$ is zero when $\mu$ and $\tau$ are zero, that is, under the null hypothesis than no SNPs have any effect.
However, when $\tau>0$, $\beta_a^{p}$ is generally positive even if~$\mu=0$, increasing quadratically with $\tau$.
Large values of $\beta_a^{p}$ with $\mu=0$ are only generated with fairly large values of $\tau$; for example,
a per-allele OR of about 1.5 is found with $\tau \approx 1.0$, meaning that the median absolute SNP OR is near 2.0.


In a brief simulation study, we validate the theoretical results in equation \eqref{logisticnormal} and Figures \ref{prevelence_curve}, \ref{mu_curve}, \ref{tau_curve}.
We simulate large cohorts varying each of the parameters ($\mu, \tau, c$ and the mean of $g_i$) and tabulate the realized disease status by $g$; Tables \ref{prospectiveVaryPrevTable} - \ref{prospectiveVaryMafTable} display the results, which are in agreement with the Figures.
The complete specification of the simulation details is found in the Supplement.

Figure \ref{sd_v_power} takes a slightly different approach and instead illustrates that the power of a t-test of allele counts (a) is non-trivial when the mean lOR is zero and the variance is non-zero (b) increases markedly with the variance of lORs.



\subsection{SNP ascertainment in case-control studies}

Having obtained the regression association for all SNPs at a given MAF, we turn to burden lORs in phase two's replication cohort with either (a) the set of SNPs detected as polymorphic (custom genotyping) or (b) novel SNPs in the same gene (sequencing).

The key to understanding our results is that the probability of passing the filter ``was the SNP polymorphic in phase one?'' depends on the odds-ratio of that SNP.
Compared to prospective designs, case-control studies are somewhat more efficient for discovering rare risk-increasing SNPs than risk-decreasing SNPs \cite{longmate_three_2010, curtin_identifying_2009, edwards_enriching_2011, edwards_enriching_2011, li_discovery_2009, yang_two-stage_2011} since frequency among the cases is an increasing function of the OR.
The set of SNPs which are polymorphic in phase one will therefore have a modestly positively deviated mean lOR compared to the set of all SNPs in the gene, and the set of SNPs \emph{not} polymorphic in phase one (and therefore eligible to be novel in phase two) will have a modestly downward deviated mean lOR.
Therefore, for these selected populations of SNPs the correct $\mu$ and $\tau$ in equation \eqref{logisticnormal} and \eqref{marginalmodel} must be modified to reflect the filtered set of SNPs.


We can make this insight more rigorous.
With trivial algebra, the probability of exposure to a minor allele of SNP $j$ among the controls and cases in a primary study is
\begin{align} 
 \textrm{Pr}\lbrace G_{ij}=1 | i\in \textrm{controls} \rbrace^{-1} &=  1 + \frac{1-f_j}{f_j} \left(1 - K +K \cdot OR_j \right)    \label{exposure_pdds} \\
\textrm{Pr}\lbrace G_{ij}=1 | i\in \textrm{cases} \rbrace^{-1} &=  1 + \frac{1-f_j}{f_j} \left( \frac{1-K}{OR_j} + K\right) \label{exposure_pdds_ca} \ ,  
\end{align}
where $f_j$ is the MAF of the $j^{th}$ SNP, $K$ the disease prevalence among homozygotes with the major allele, and $OR_j=e^{\gamma_j}$ the odds-ratio.
When both $f_j <<1$ and $K<<1$ the expressions can be approximated by
\begin{align} 
 \textrm{Pr}\lbrace G_{ij}=1 | i\in \textrm{controls} \rbrace & \approx  \left(1 + \frac{1+K \cdot OR_j}{f_j}\right)^{-1} \approx f_j     \\
\textrm{Pr}\lbrace G_{ij}=1 | i\in \textrm{cases} \rbrace & \approx  \left(1 + \frac{1-K}{OR_j \cdot f_j}\right)^{-1} \approx  OR_j \cdot f_j \ .  
\end{align}
With the same assumption that the MAF and disease prevalence are low, the probability that a given SNP $j$ is observed at least once (i.e., is \emph{ascertained} as polymorphic) in either the $N$ cases or $N$ controls of phase one can be obtained using a Poisson approximation: 
\begin{equation}\label{missed-pr}
 \pr \left\{ \sum_i G_{ij} > 0 \right\} \approx 1-\exp(-{f_j\ N}(1+OR_j)).
\end{equation}

\begin{figure}[ht]
\begin{center}
\centerline{\includegraphics[width=6in]{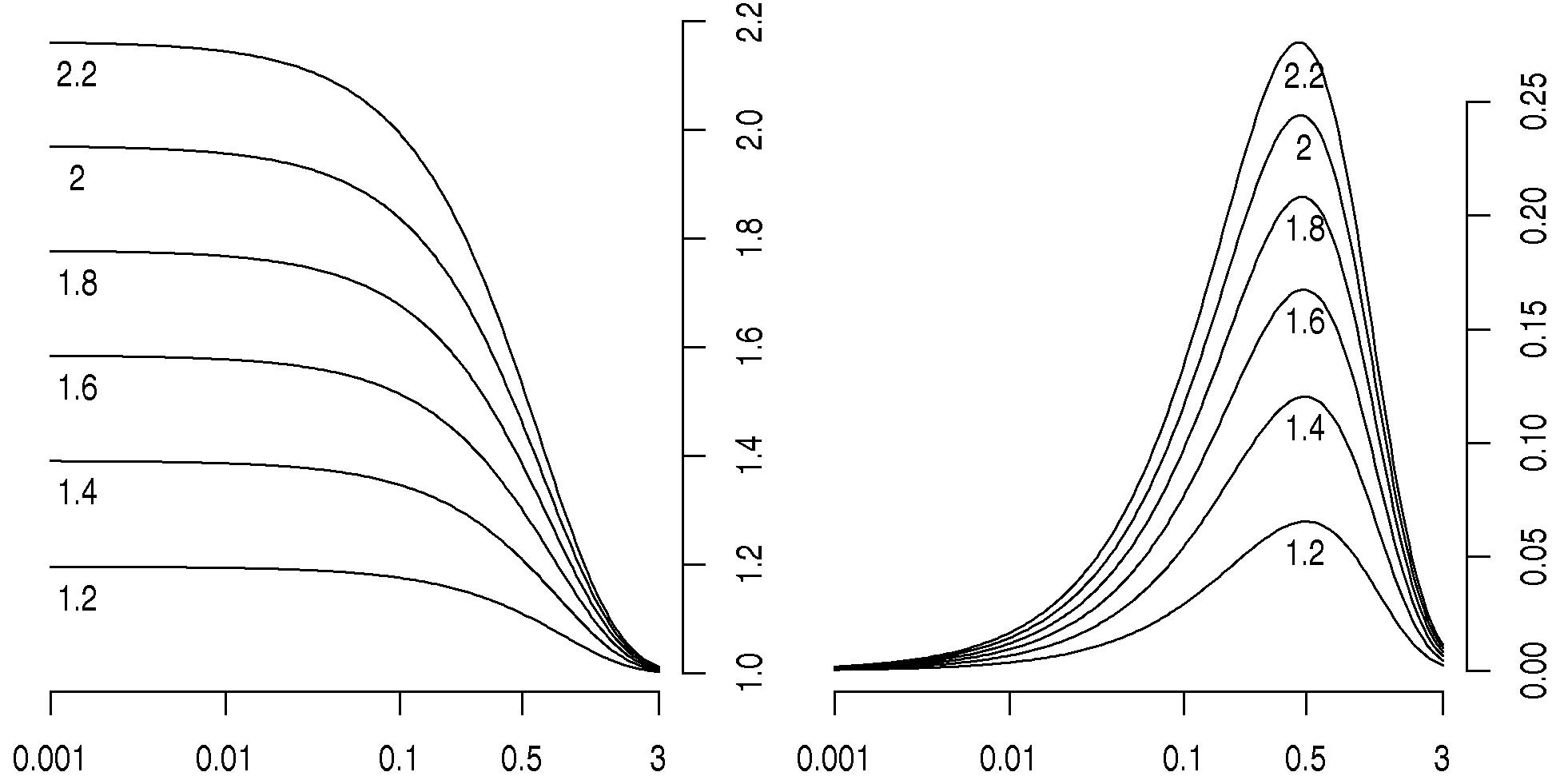}}
\end{center}
\caption[Sampling probabilities by OR and MAF]{ {\bf Sampling probabilities by OR and MAF}
x-axis: $N \cdot f$. Left panel y-axis: ratio of sampling probability for a SNP at the implied MAF with OR given by the label to the sampling probability under the inverse of that OR. Right panel: corresponding difference in sampling probabilities. N=100, prevalence=1\%.
}
\label{multibias}
\end{figure}

There are three important things to note about formula \eqref{missed-pr}.
First, risk-associated alleles are enriched into the case-sample, but the enrichment of disease-preventing alleles into the control-sample is negligible in a rare-disease setting.
Examining Equation \eqref{exposure_pdds}, one can see that decreasing the odds ratio from one to zero only slightly changes the probability of exposure in controls when $1-K >> K$; however, the probability of exposure in cases goes to zero almost linearly.
Figure \ref{multibias} displays the ratio and the difference of polymorphic probabilities for risk-increasing versus risk-decreasing alleles with the same magnitude log-odds ratio; the risk-increasing allele is more likely to be sampled for all MAFs and magnitudes of lOR.
%

\begin{figure}[!ht]
\begin{center}
\includegraphics[width=6in]{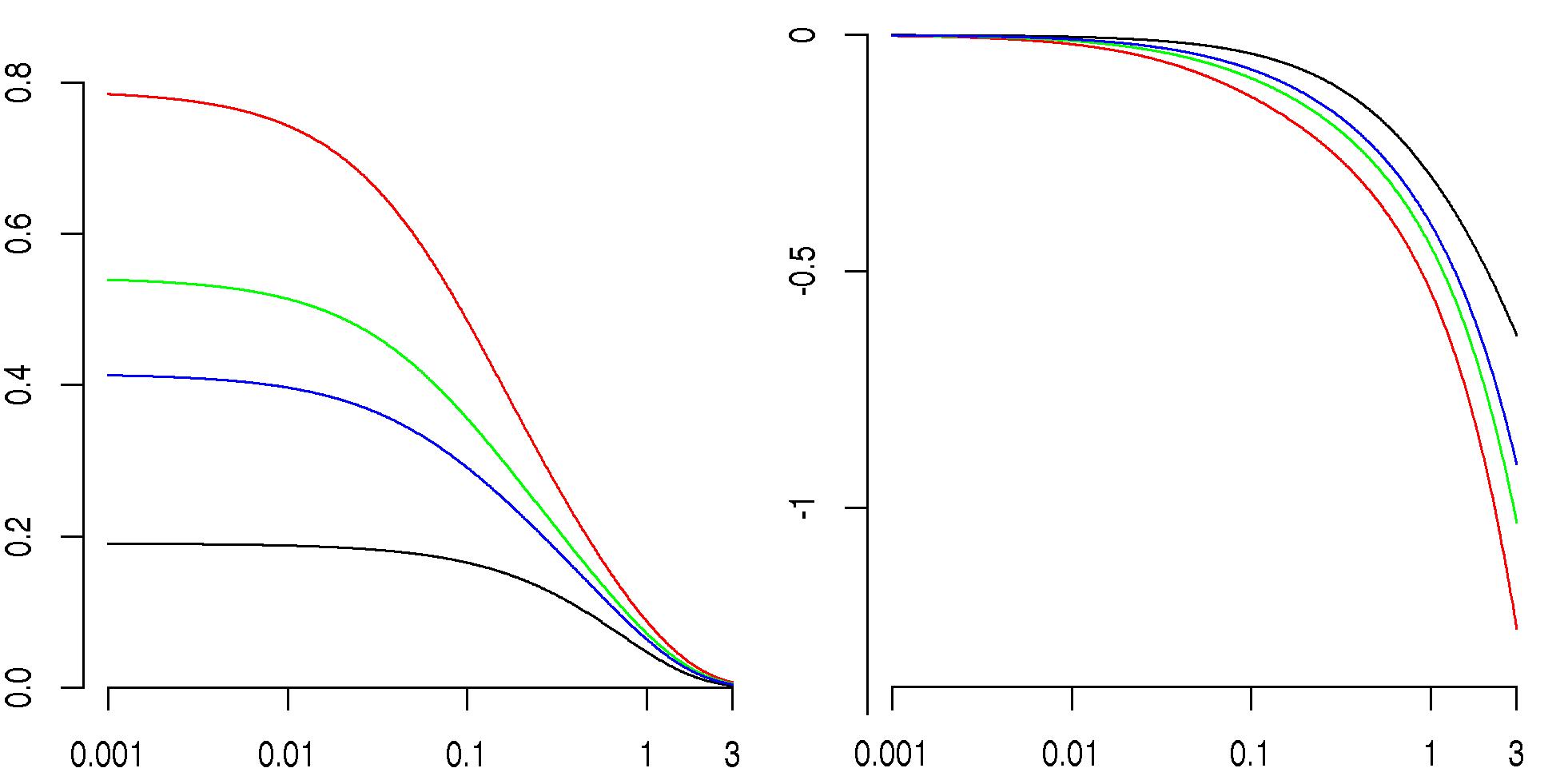}
\end{center}
\caption[Expected lOR by sampling status]{ {\bf $\mu_{p}$ (left) and $\mu_{r}$  (right) versus MAF by tail behavior of distribution of SNP lORs.}  x-axis: $N \cdot f$. y-axis: mean lOR. Source log-odds ratio distributed N(0,$0.6^2$) (black), Student's t distributions with 1, 2, 3 degrees of freedom (green, red, blue) truncated at $\pm 4$ with same 20\% and 80\% quantile as N(0,$0.6^2$). N=100, prevalence=1\%.
}
\label{expected_rare}
\end{figure}

Figure \ref{multibias} displays the effect of retrospective sampling on fixed lORs, but it is more natural to consider lORs which come from a continuous distribution.
Figure \ref{expected_rare} shows the expected value of previously observed and novel lORs in a replication or prediction sample (that is, $\mu_{p}$ and $\mu_{r}$) for several members of the truncated Gaussian and Student's t distribution family. 
The right panel of Figure \ref{expected_rare} displays an interesting consequence of the above facts.
Risk increasing alleles are increased in the case sample and total likelihood of being observed and protective alleles are depleted; failing to discover a SNP with  MAF $> 1/N$ suggests that this SNP has low frequency among cases and is protective.
However, as MAF increases the rate of non-polymorphic SNPs becomes very low, and this feature is of little practical importance.

Distributions with heavier tails than a Gaussian exhibit a higher bias and an exaggerated effect at very low MAF.
The cumulative distributions plots of the lORs under these scenarios are displayed in Figure \ref{density_plots}, which shows that they are very similar in the region of zero and primarily differ in the likelihood of large uncommon lORs which ``rescue'' occasional low-MAF SNPs into the case sample.

Second, due to the exponential form of formula \eqref{missed-pr}, the above bias is meaningful only for a narrow range of MAF and peaks for SNPs with MAF $\approx 1/2N$.
Common minor alleles are virtually assured of being polymorphic in phase one, so the impact of differential ascertainment becomes of negligible magnitude.
As shown in the right panel of Figure \ref{multibias} the difference between sampling probabilities also goes to zero for very low allele frequencies; even a quite large OR does not include many very rare SNPs in the observed set. 
%
%
In Figure \ref{expected_var}, the variance of polymorphic SNP lORs varies with MAF; however, the change is small in the Gaussian case and only notable for t distributions which allow large SNP effects and only for comparatively rare SNPs.
Previously observed SNPs (left panel Figure \ref{expected_var}) have elevated variances for low MAF, while novel SNPs have a small nadir in the variance around $f \approx 1/2N$ for the examined t distributions.
Intuitively, case-control sampling allows rare SNPs with quite large effects to be sampled into the cases, increasing the variability of observed lORs.

Third, the polymorphic probability in formula \eqref{missed-pr} depends on the sample size, rather than just the minor allele frequency.
Above we argued that non-representative sampling can only be a problem for a particular range of low to rare frequencies; however, that range is a function of the sample size.
This creates a differential bias between the SNP sets observed in studies of different sample sizes and ensures that as long as there are low MAF SNPs the above bias will exist regardless of the sample size.
In Figure \ref{shift_multibias} we recreate Figure \ref{expected_rare} but with an increased sample size, which shifts the location of the curve but leaves the shape the same.
Finally, a feature lost in approximation \eqref{missed-pr} is the dependence on baseline disease prevalence.
Examining the probabilities in Equation \ref{exposure_pdds_ca}, as the prevalence goes to zero the enrichment of risk increasing SNPs in cases is maximized.
This is also illustrated in Figure \ref{prevelence_curve}.

\subsection{Allele count lORs in phase two}
\label{ascertainment_bias_section}
Exploiting the foregoing results, the left-hand panels of Figures \ref{prevelence_curve}, \ref{mu_curve}, and \ref{tau_curve} display the mean lOR in the phase two study for previously polymorphic ($\mu_p$) and novel ($\mu_r$) SNPs derived using equations \eqref{exposure_pdds}-\eqref{exposure_pdds_ca}. 
Using the results from Section \ref{prim-allele-count-effects} (equations \eqref{logisticnormal} - \eqref{marginalmodel}), the right-hand panels of those figures then display the allele count lOR expected for each group of SNPs in phase two.

The left panel of Figure \ref{prevelence_curve} shows that previously sampled and novel SNPs have notably different mean lORs, $\mu_p > \mu_r$ and that this difference increases linearly as the prevalence becomes small.  
The right panel similarly shows that the burden lOR is markedly different between previously polymorphic and novel SNPs for rare diseases, and that the difference decreases greatly as the prevalence goes from 1\% to 15\%.  
Figure \ref{mu_curve} takes the same strategy varying the mean SNP effect $\mu$.  
The difference in true means of lORs and burden lORs between previously polymorphic and novel SNPs does change, but not much within the plausible range of $\mu$.  
Figure \ref{tau_curve} again shows the variance of SNP lORs to be a primary determinant of the difference in $\beta$.  
The difference in all $\beta$'s is zero at $\tau=0$ (the null), but increases quickly with $\tau$, especially beyond 0.5.
Interestingly, the change in $\beta$'s with $\tau$ does not quite mirror the change in $\mu$'s with the burden lOR for SNPs novel to the replication nearly constant around zero despite large negative changes in the mean lOR of those SNPs.

To illustrate the variation within samples and the effect when SNP lORs are fixed, a scatter plot of $\hat{\beta^r_p}$ and $\hat{\beta^r_r}$ versus $\hat{\beta^r_a}$ from simulations (described below) is shown in \ref{beta_scatter}.
The estimated effects are always $\hat{\beta_{r}^{r}} < \hat{\beta_{a}^{r}} < \hat{\beta_{p}^{r}}$, and the three highly correlated across realizations of $\gamma$.

We validate these calculations in a second brief simulation study. 
In each simulation, we draw a balanced case-control sample from a large population as well as an independent prospective replication cohort; details of the simulation setup are found in the Supplement and the results in 
Tables \ref{retroVaryPrevTable}, \ref{retroVaryTauTable}, \ref{retroVaryMuTable}, \ref{retroVaryMafTable}, \ref{retroVaryMafNJointTable}, \ref{retroVarySampSizeTable}, and \ref{retroVarySampSizeImportantTable}.
The tables recapitulate the claims above along with some new findings.
In specific, the correct prediction for previously observed SNPs is that they are much more risk increasing than  novel SNPs.
As the mean or variance of SNP effects increases, the difference becomes exaggerated (Tables \ref{retroVaryMuTable} and \ref{retroVaryTauTable}).
With MAFs which are small compared to the sample size, previously observed SNPs are more risk-increasing and at larger MAFs novel SNPs are very protective (Table \ref{retroVaryMafTable} and \ref{retroVarySampSizeTable}).
The sample size determines the MAF at which the difference in SNP classes is largest, but not the maximum magnitude of the effect (Table \ref{retroVarySampSizeTable} and \ref{retroVaryMafNJointTable}).


\section{Discussion}
Our findings can be summarized around two central points.
First, we clarify the factors which affect the expected lOR of allele-count burdens, and find that they are not an intuitive reflection of the SNP-specific model which generates the data.
The allele-count to phenotype correlation is determined by the variance of SNP effects and disease prevalence as well as the mean lOR of SNPs.
We focus on the case where  the mean log-odds ratio is zero and provide examples in hypothesis tests (Figure \ref{sd_v_power}) and prediction (Figure \ref{tau_curve}) where an investigator could correctly claim than possessing more minor alleles increased risk of disease.
The differing interpretation of the mean effect of increasing allele-counts and the mean of SNP lORs is a product of the asymmetric baseline risk of being affected with disease and the nonlinear nature of logistic models of risk.
An increased variance of SNP lORs will change the disease risk for those individuals carrying rare alleles in both directions; however, there are many more undiseased individuals at risk of becoming diseased than vice versa, so increasing positive and negative lORs do not cancel out in their effects on the population.

Second, we show that replication experiments which selectively genotype rare SNPs previously observed in a case-control study will tend to find a \textit{larger} allele count lOR than in the primary study.
The complementary set of SNPs (those not seen in the primary case-control sample and novel to the replication experiment) tend to increase risk less than average or even decrease it; further sequencing to find more rare SNPs will experience a unique ``winner's curse'' as shown in Figure \ref{tau_curve}.
A different application of the same finding is that novel alleles in highly associated genes in individuals sequenced for medical prediction tend to carry \textit{much less risk} than the SNPs in a case-control discovery study.
This can be understood by considering the impact of SNP lOR on the likelihood of a SNP being polymorphic in a case-control study.
Risk-decreasing minor alleles have an decreased frequency among cases compared to their population MAF and are more likely than SNPs which increase risk to be observed zero times in the study; these low-impact and negatively associated SNPs have been filtered out.
Further data collection which uses a case-control study to select SNPs to genotype is therefore partially conditioning on the effect size of those SNPs.
This difference between previously observed and novel SNPs is mostly observed at MAF near $1/2N$, and therefore predictions based on experiments of different size can appear inconsistent.

We have represented SNP lORs using a continuous distribution.
This is not a prior uncertainty like in a Bayesian analysis.
From gene to gene, we can expect the \textit{realized} distribution of SNP effects to vary even if there is no fundamental difference in the propensity for SNPs in that gene to affect the trait.
Similarly, within a gene we foresee substantial variation in realized rare SNP lORs across human populations with SNPs private to their ancestry and by sampling variation within human populations (for example between the initial association study and a replication cohort).


%


We have represented SNP lORs using a continuous distribution.
This is not a prior uncertainty like in a Bayesian analysis.
From gene to gene, we can expect the \textit{realized} distribution of SNP effects to vary even if there is no fundamental difference in the propensity for SNPs in that gene to affect the trait.
Similarly, within a gene we foresee substantial variation in realized rare SNP lORs across human populations with SNPs private to their ancestry and by sampling variation within human populations (for example between the initial association study and a replication cohort).
These findings are preceded by several related ideas in the biostatistics literature.
In the terms of the causal inference literature, predictors in logistic model are not ``collapsible'' \cite{greenland_confounding_1999}.
While it has been observed that a SNP specific (random effect) model is richer and answers additional questions, the parameters can have an awkward interpretation and extrapolation to new data, as we have observed.
This distinction originated in biometrics modeling longitudinal data \cite{zeger_models_1988}, but has been reiterated in many contexts \cite{crouchley_comparison_1999, hu_comparison_1998, hubbard_gee_2010, subramanian_modeling_2010}.

Coming from different perspectives, many previous authors have noted that compared to prospective designs, case-control studies are somewhat more efficient for discovering rare risk-increasing SNPs \cite{longmate_three_2010, curtin_identifying_2009, edwards_enriching_2011, edwards_enriching_2011, li_discovery_2009, yang_two-stage_2011}, generally taken as an argument increasing the power of case-control study.
There is also a related literature on ascertainment-adjustment to variance components based analysis of family data, which stems from a similar problem \cite{burton_ascertainment_2000, glidden_ascertainment_2002, bowden_two-stage_2007, ma_ascertainment_2007, noh_robust_2005, epstein_ascertainment-adjusted_2002, sung_genetic_2005, burton_covariance_2005, burton_correcting_2003}.
In family designs the relatively simple structure of ascertainment makes correction feasible.
Similarly, SNP-panel and small sequencing study based inferences on site frequency spectra and other population genetic quantities have been examined for ascertainment bias \cite{nielsen_reconstituting_2004, nielsen_correcting_2003, clark_ascertainment_2005}.



An important limitation to our results is that the calculations only produce substantial bias in the presence of fairly strong SNP effects (Figures \ref{tau_curve} and \ref{expected_rare}), which might be unrealistic.
However, the supposed existence of rare SNPs with large effects is a major rationale for performing sequencing-based association studies \cite{manolio_finding_2009}, and that the failure to detect such variants in GWAS does not exclude their existence.
We suspect that a distribution like the truncated Student's t used in Figure \ref{expected_rare} whose distribution is plotted in Figure \ref{density_plots} represents a reasonable belief about log-odds ratios in causal genes; most are very nearly zero with a few outlying strong effects.

A second limitation of these findings is that they ignore external data on SNP MAF.
With numerous sequencing projects published and in development, ``discovery'' of SNPs during future case-control projects may be very limited, and the information on MAF provided by external data should stabilize per-SNP inferences about ORs.
However, because of the human SNP site-frequency spectrum, additional sequencing is predicted to continue discovering new SNPs for quite some time \cite{coventry_deep_2010, keinan_recent_2012}.
Additionally, existing panels such as 1000 genomes do not provide much MAF information for SNPs unique to unincluded populations such as Native Americans, and population isolates of interest to particular diseases can have markedly different MAFs due to drift as well as private SNPs.
Finally, many non-human organisms do not yet have extensive sequencing catalogs, but their genetics are of scientific and applied importance and investigated by similar scientific methods.

Third, we have not presented simulations or calculations where MAFs are drawn from a continuous distribution as they would be in real data.
Because the difference between primarily polymorphic and novel in replication SNPs varies with MAF, we feel that such scenarios add an unnecessary level of complication to understanding the phenomenon without creating additional insight.
Approximate results for a mixture of MAFs can be obtained by a weighted average of the results for point distributions of MAF.
Uncommon SNPs (especially singletons) observed in a real sequencing study are a mixture of those with MAF near $1/2N$ and lucky representatives of the large pool of rarer SNPs.
Fortunately, as seen in Figures \ref{multibias} and \ref{expected_rare} the relative bias stabilizes as the MAF becomes small.
The impact of MAF is still important to keep in mind because comparisons across genes with different SNP frequency spectra will experience differential bias.
Similarly, there will be differential bias when comparing human populations with different demographic history and therefore SNP frequency distributions  \cite{gutenkunst_inferring_2009}.

We have reserved study of methods for correction of the ascertainment effects of section \ref{ascertainment_bias_section} for future work.
Calculating the bias for known distributions of MAF and SNP effects is trivial; however, applying this correction in practice is challenging for several reasons.
First, as we showed in Figure \ref{expected_rare} the result depends on the tails of the distribution of lORs, which can be difficult to accurately estimate.
Second, the observed data are minimally informative for estimation of the MAF spectrum around $1/2N$ and especially for extrapolation to lower MAF.
External information in the form of demographic models and previously sequenced cohorts of the same ethnicity will be crucial to developing accurate estimates and can not be spoken to in much generality.
Third, the joint distribution of MAF and SNP lORs matters.
We have performed calculations only where the two are independent, but if the two are correlated, as models of phenotype selection imply \cite{king_evolutionary_2010}, then a more complex procedure is required which estimates the distribution of SNP effects for low MAFs and  extrapolates to very low MAFs. 



\section*{Acknowledgments}
We are grateful to Nancy Cox and Lin Chen for comments on a draft of the paper.
We would like to thank Jonathan Pritchard for feedback on the project.

\clearpage
\bibliography{new_asc}


\section*{Supplement}

\setcounter{figure}{0}
\setcounter{table}{0}
\makeatletter 
\renewcommand{\thefigure}{S\@arabic\c@figure} 
\renewcommand{\thetable}{S\@arabic\c@table} 
\makeatother

\begin{figure}[!ht]
\begin{center}
\centerline{\includegraphics[width=3.27in]{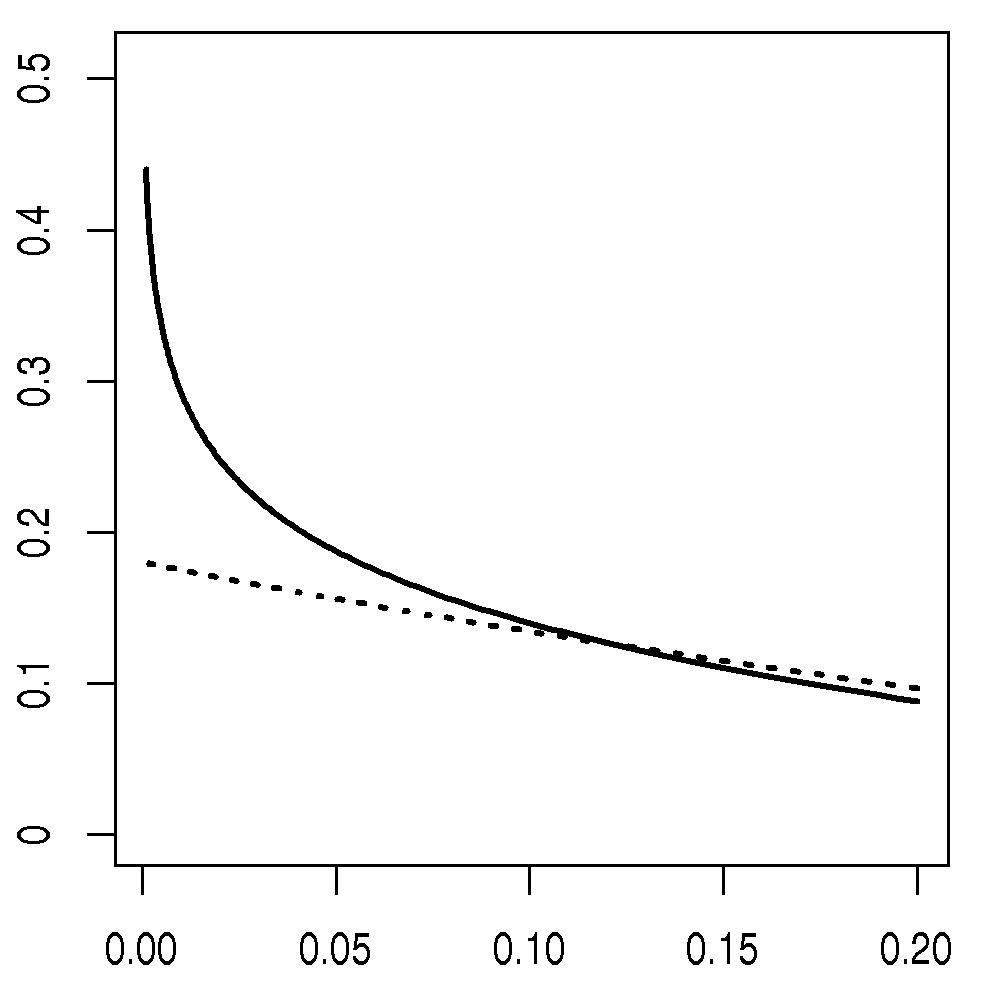}}
\end{center}
\caption{
{\bf Comparison of formulas \eqref{marginalmodel} and \eqref{logisticnormal}. } Disease prevalence among those with no minor alleles (x-axis) versus burden lOR from equation \eqref{logisticnormal} (dotted) and formula \eqref{marginalmodel} (solid). lOR of 1 SNP versus 0 SNPs. SD of SNP lORs 0.6, mean of lORs 0, MAF 0.005. Low limit of plot = 0.001.
}
\label{normal_approx_bad}
\end{figure}

\begin{figure}[!ht]
\begin{center}
\includegraphics[width=6in]{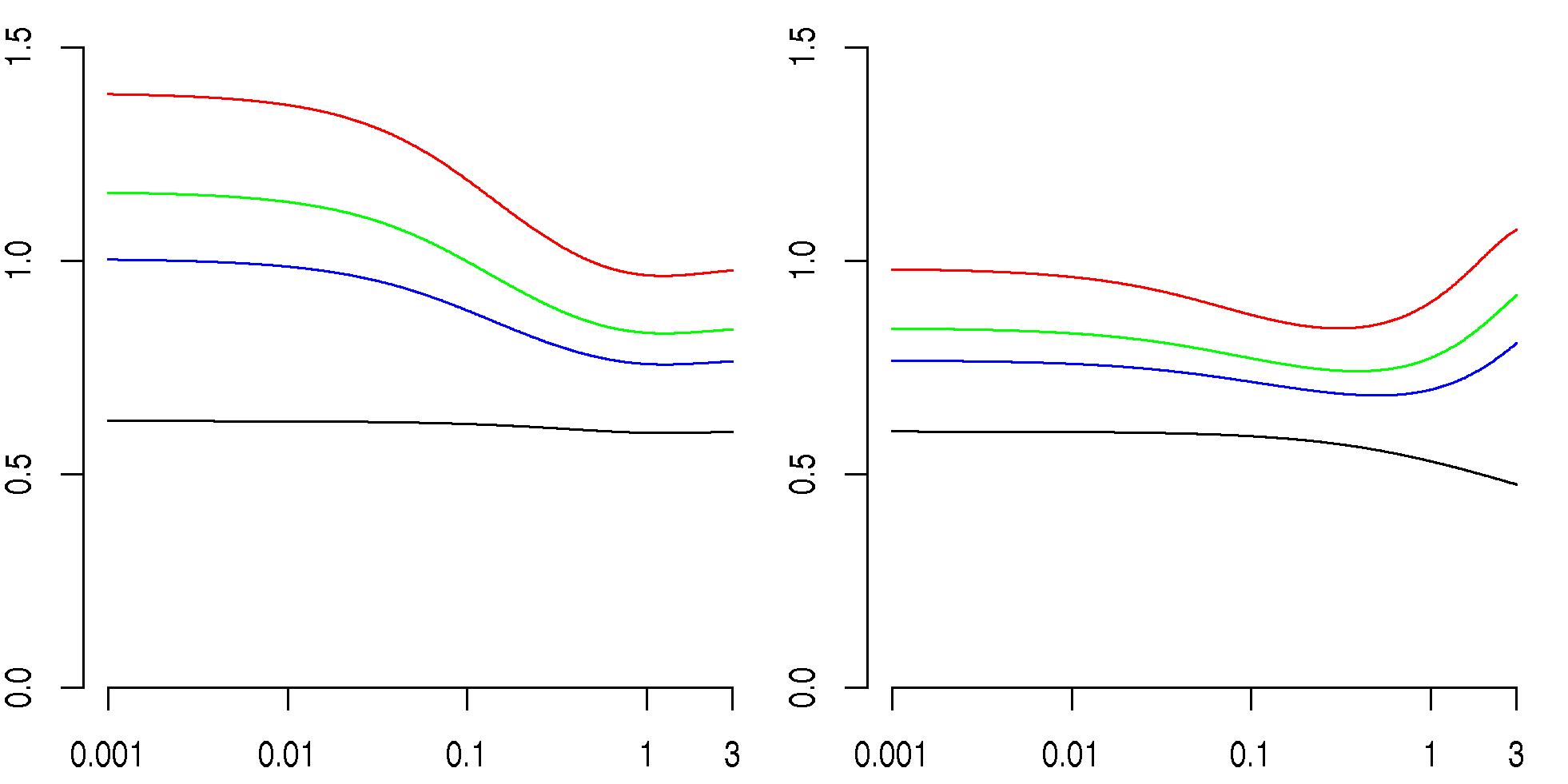}
\end{center}
\caption[Standard deviation of lOR by sampling status]{ {\bf $\tau_{p}$ (left) and $\tau_{r}$ (right) versus MAF by tail behavior of distribution of SNP lORs.}  Settings identical to Figure \ref{expected_rare}.
}
\label{expected_var}
\end{figure}

\begin{figure}[!ht]
\begin{center}
\includegraphics[width=5in]{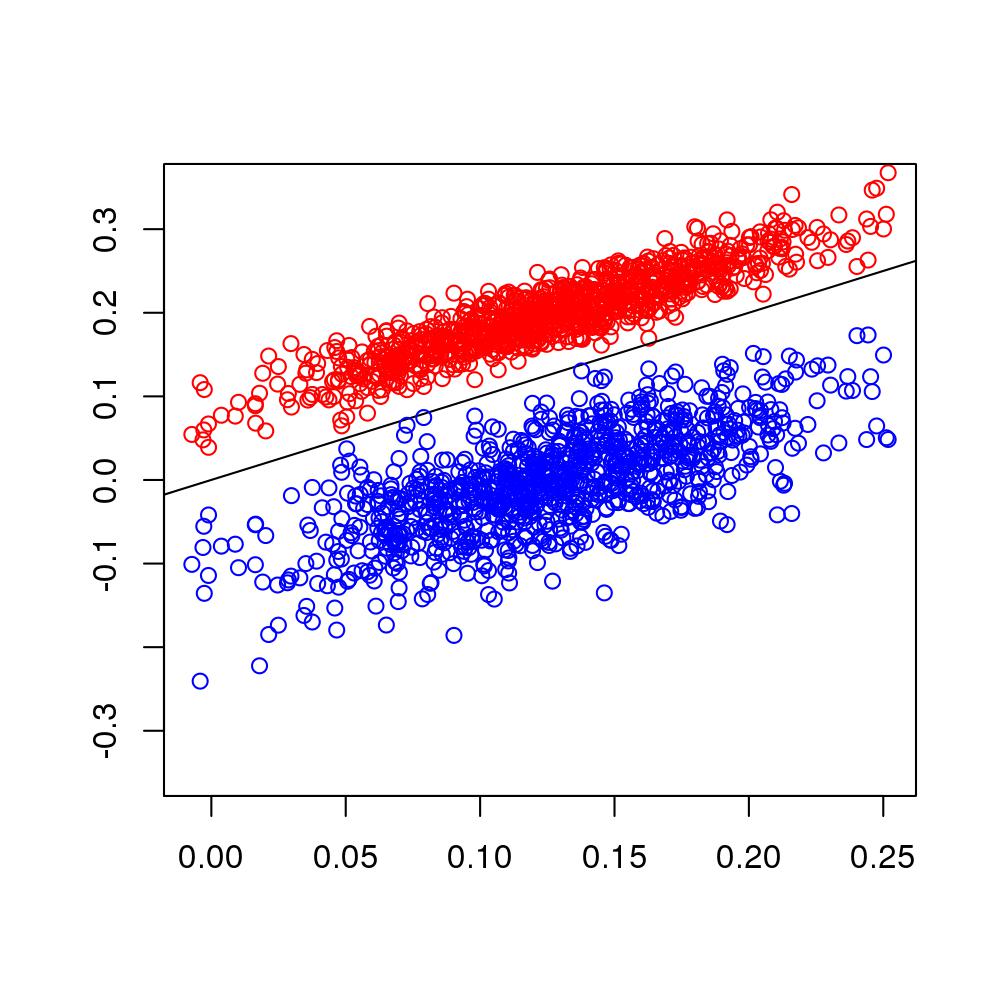}
\end{center}
\caption{ $\hat{\beta^r_p}$ (red) and $\hat{\beta^r_r}$ (blue) vs $\hat{\beta^r_a}$ (x-axis position). Black line identity. Prevalence 1\%, MAF .005, $mu=0$, $\tau=0.6$, N=100, 35 SNPs, total population size $10^7$, prediction cohort N = $10^6$.
}
\label{beta_scatter}
\end{figure}

\subsection{Simulation settings}
\begin{figure}[!ht]
\begin{center}
\centerline{\includegraphics[width=3.27in]{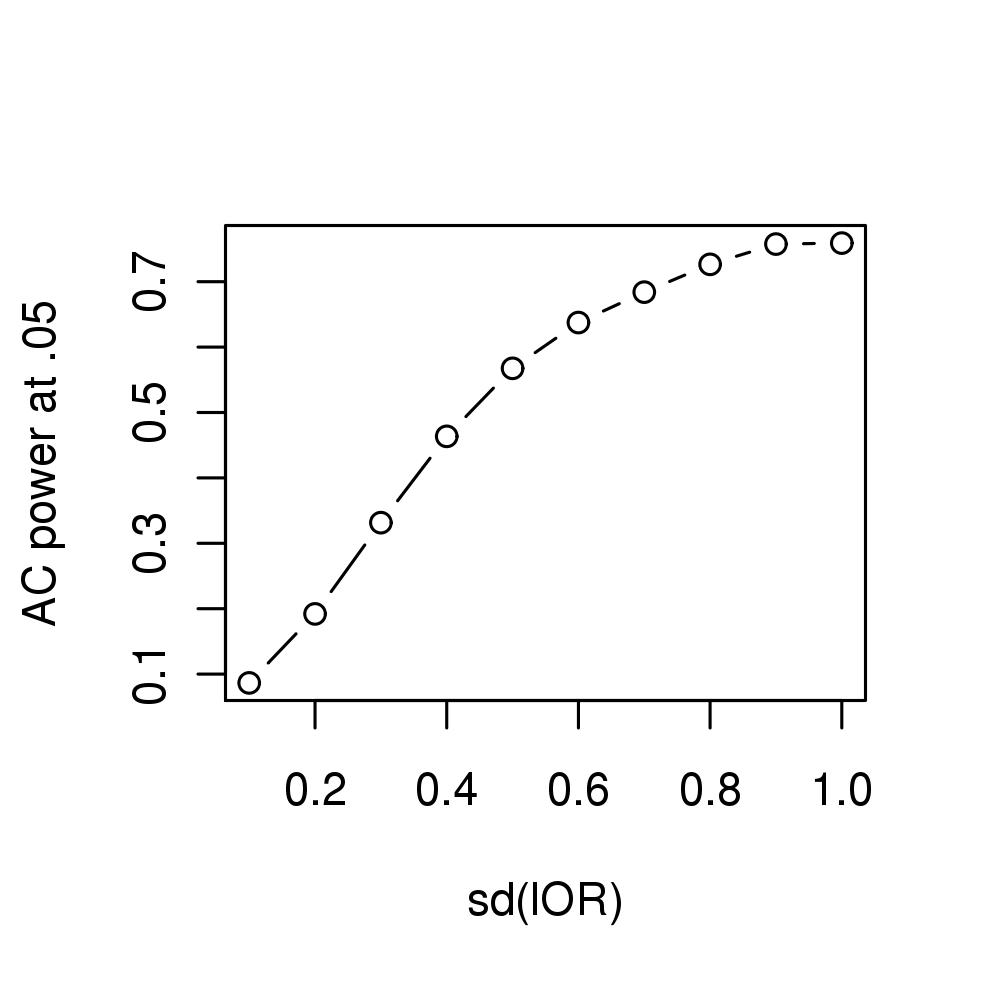}}
\end{center}
\caption{
{Standard deviation of log-odds ratios (x-axis) and simulated power of t-test of allele count (y-axis) in case-control design when the log-odds ratios are Gaussian with mean zero, N=500, prevalence .05, 40 SNPs, MAF drawn from a beta distribution with mean .12 and SD .02, 2000 replicates.}
}
\label{sd_v_power}
\end{figure}

For Figure \ref{sd_v_power} we simulated 40 SNPs with minor allele frequencies drawn from a beta distribution mean .12 standard deviation .02 in a population of 1 million individuals with a baseline disease prevalence of 5\%, drawing a sample of 500 cases 500 controls for each of the depicted values of the standard deviation of log-odds ratios.  
Log odds-ratios were always mean zero IID normal. We then conducted a t-test of the number of minor alleles in cases and controls; essentially identical results were obtained for rank-tests and generalized linear models.  
2000 replicates for each point in each figure were produced, each repeating the entire procedure.

Tables \ref{prospectiveVaryPrevTable}-\ref{retroVarySampSizeImportantTable} use a Gaussian distribution for lORs and a large sample prospective ($10^6$) generated by the logistic model \eqref{ss-m1} with all SNPs uncorrelated.
2000 replicates are drawn for each Table.
Table \ref{prospectiveVaryPrevTable} varies the disease prevalence in rows and holds the SD of SNP effects at 0.6, the mean SNP effect at 0, the MAF at 0.01, and the number of SNPs at 50. 
Table \ref{prospectiveVaryTauTable} varies the SD of SNP effects in rows and holds the the mean SNP effect at 0, the disease prevalence at 5\%, the MAF at 0.01, and the number of SNPs at 50. 
Table \ref{prospectiveVaryMuTable} varies the mean of SNP effects in rows and holds the SD of SNP effects at 0.7, the disease prevalence at 5\%, the MAF at 0.01, and the number of SNPs at 50. 
Table \ref{prospectiveVaryMafTable} varies the SNP MAF in rows and holds the SD of SNP effects at 0.7, the mean SNP effect at 0, the disease prevalence at 5\%, and the number of SNPs at 50. 
Tables for retrospective simulation draw 100 cases and 100 controls from a source population of $10^6$ and compare the results in $10^6$ future independent individuals.
We display only the contrast for having no minor alleles versus 1 minor allele for all alleles combined, novel in follow-up, and previously discovered SNPs; patterns in the impact of the number of alleles are similar to the prospective case (not shown).
Table \ref{retroVaryPrevTable} varies the disease prevalence in rows and holds the SD of SNP effects at 0.6, the mean SNP effect at 0, the MAF at 0.01, and the number of SNPs at 50. 
Table \ref{retroVaryTauTable} varies the SD of SNP effects in rows and holds the the mean SNP effect at 0, the disease prevalence at 5\%, the MAF at 0.01, and the number of SNPs at 50. 
Table \ref{retroVaryMuTable} varies the mean of SNP effects in rows and holds the SD of SNP effects at 0.7, the disease prevalence at 5\%, the MAF at 0.01, and the number of SNPs at 50. 
Table \ref{retroVaryMafTable} varies the SNP MAF in rows and holds the SD of SNP effects at 0.7, the mean SNP effect at 0, the disease prevalence at 5\%, and the number of SNPs at 50. 
Table \ref{retroVaryMafNJointTable} jointly varies the number of SNPs and the MAF to keep the average number of minor alleles per person at 0.5; it holds the SD of SNP effects at 0.7, the mean SNP effect at 0, and the disease prevalence at 5\%.
Table \ref{retroVarySampSizeTable} varies the case-control sample size in rows and holds the SD of SNP effects at 0.7, the mean SNP effect at 0, the disease prevalence at 5\%, the MAF at 0.01, and the number of SNPs at 50.
Table \ref{retroVarySampSizeImportantTable} jointly varies the MAF and the sample size in rows to keep the expected number of appearances for each SNP constant at 0.5; it holds the SD of SNP effects at 0.7, the mean SNP effect at 0, the disease prevalence at 5\%, and the number of SNPs at 50.


\begin{table}[ht]
\begin{center}
\begin{tabular}{c|cc|cc|cc|}
  \hline
 & 1 &   & 2 &   & 3 &   \\ 
  \hline
0.2 & 0.09 & 0.09 & 0.16 & 0.17 & 0.22 & 0.23 \\ 
  0.1 & 0.14 & 0.13 & 0.26 & 0.24 & 0.35 & 0.33 \\ 
  0.05 & 0.19 & 0.15 & 0.34 & 0.28 & 0.48 & 0.40 \\ 
  0.01 & 0.29 & 0.18 & 0.54 & 0.33 & 0.74 & 0.48 \\ 
   \hline
\end{tabular}
\caption{lOR of outcome by number of derived alleles (log odds of disease given g = column index - log odds given g = 0)  from formula \eqref{marginalmodel} (left number) and simulation (right number) varying prevalence of disease (first column). Prospective sampling design..}
\label{prospectiveVaryPrevTable}
\end{center}
\end{table}
\begin{table}[ht]
\begin{center}
\begin{tabular}{c|cc|cc|cc|}
  \hline
 & 1 &   & 2 &   & 3 &   \\ 
  \hline
0.25 & 0.04 & 0.03 & 0.07 & 0.05 & 0.10 & 0.07 \\ 
  0.5 & 0.13 & 0.10 & 0.25 & 0.20 & 0.35 & 0.28 \\ 
  0.75 & 0.28 & 0.23 & 0.49 & 0.42 & 0.66 & 0.57 \\ 
  1 & 0.45 & 0.38 & 0.74 & 0.65 & 0.95 & 0.85 \\ 
  1.25 & 0.62 & 0.55 & 0.97 & 0.89 & 1.19 & 1.11 \\ 
   \hline
\end{tabular}
\caption{lOR of outcome by number of derived alleles (log odds of disease given g = column index - log odds given g = 0)  from formula \eqref{marginalmodel} (left number) and simulation (right number)  varying SD of SNP effects (first column). Prospective sampling design.}
\label{prospectiveVaryTauTable}
\end{center}
\end{table}
\begin{table}[ht]
\begin{center}
\begin{tabular}{c|cc|cc|cc|}
  \hline
 & 1 &   & 2 &   & 3 &   \\ 
  \hline
-0.47 & -0.18 & -0.25 & -0.36 & -0.52 & -0.53 & -0.79 \\ 
  0 & 0.25 & 0.21 & 0.44 & 0.38 & 0.60 & 0.52 \\ 
  0.26 & 0.49 & 0.46 & 0.89 & 0.86 & 1.22 & 1.21 \\ 
  0.47 & 0.68 & 0.66 & 1.24 & 1.24 & 1.72 & 1.74 \\ 
  0.69 & 0.88 & 0.87 & 1.62 & 1.62 & 2.26 & 2.25 \\ 
  0.92 & 1.09 & 1.08 & 2.00 & 2.01 & 2.79 & 2.80 \\ 
   \hline
\end{tabular}
\caption{lOR of outcome by number of derived alleles (log odds of disease given g = column index - log odds given g = 0)  from formula \eqref{marginalmodel} (left number) and simulation (right number)  varying mean of SNP effects (first column). Prospective sampling design.}
\label{prospectiveVaryMuTable}
\end{center}
\end{table}
\begin{table}[ht]
\begin{center}
\begin{tabular}{c|cc|cc|cc|}
  \hline
 & 1 &   & 2 &   & 3 &   \\ 
  \hline
0.005 & 0.25 & 0.20 & 0.44 & 0.37 & 0.60 & 0.51 \\ 
  0.01 & 0.25 & 0.19 & 0.44 & 0.36 & 0.60 & 0.49 \\ 
  0.02 & 0.25 & 0.21 & 0.44 & 0.38 & 0.60 & 0.52 \\ 
  0.03 & 0.25 & 0.20 & 0.44 & 0.37 & 0.60 & 0.51 \\ 
  0.05 & 0.25 & 0.20 & 0.44 & 0.37 & 0.60 & 0.51 \\ 
   \hline
\end{tabular}
\caption{lOR of outcome by number of derived alleles (log odds of disease given g = column index - log odds given g = 0)  from formula \eqref{marginalmodel} (left number) and simulation (right number)  varying SNP MAF (first column). Prospective sampling design.}
\label{prospectiveVaryMafTable}
\end{center}
\end{table}

\begin{figure}[!ht]
\begin{center}
\centerline{\includegraphics[width=6in]{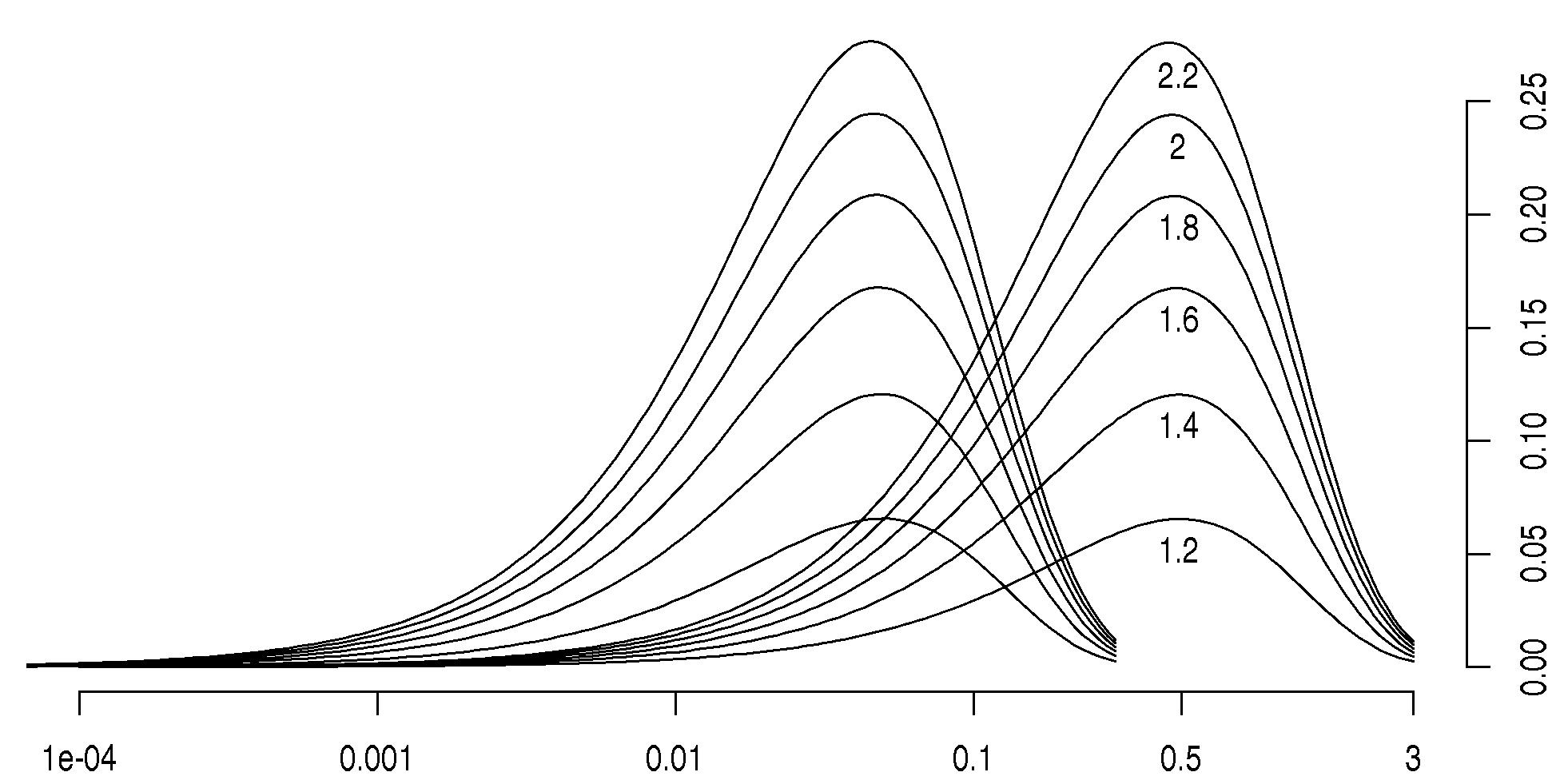}}
\end{center}
\caption{
{\bf Effect of increasing sample size on ascertainment bias.} Identical to right panel of figure \ref{multibias} with new lines displaced to the left the result with N=1000 instead of N=100. X-axis = $100\cdot f$.
}
\label{shift_multibias}
\end{figure}

%
%

\begin{figure}[!ht]
\begin{center}
\centerline{\includegraphics[width=3.27in]{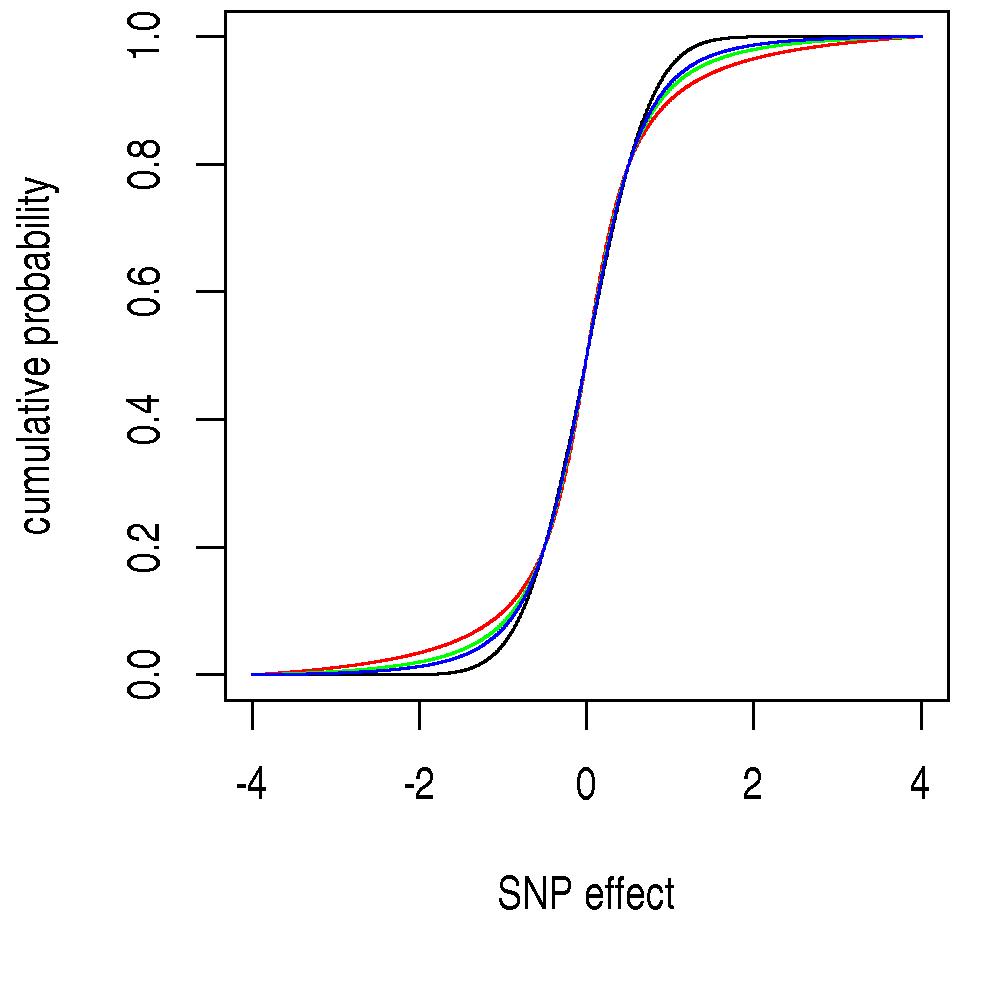}}
\end{center}
\caption{
{\bf Cumulative distribution plots for SNP log-odds ratios used in Figure \ref{expected_rare}.} N(0,$.6^2$) (black), Student's t distributions with 1, 2, 3 degrees of freedom (green, red, blue) with same inner 80\% quantile as N(0,$.6^2$).
}
\label{density_plots}
\end{figure}

\begin{table}[ht]
\begin{center}
\begin{tabular}{c|cc|cc|cc|}
  \hline
 & all &   & novel &   & old &   \\ 
  \hline
0.2 & 0.10 & 0.10 & -0.01 & -0.00 & 0.15 & 0.16 \\ 
  0.1 & 0.13 & 0.14 & -0.02 & -0.00 & 0.21 & 0.22 \\ 
  0.05 & 0.16 & 0.16 & -0.02 & -0.00 & 0.24 & 0.25 \\ 
  0.01 & 0.17 & 0.18 & -0.03 & -0.00 & 0.27 & 0.28 \\ 
  0.001 & 0.18 & 0.19 & -0.03 & -0.00 & 0.27 & 0.29 \\ 
  1e-04 & 0.18 & 0.19 & -0.03 & -0.00 & 0.27 & 0.29 \\ 
   \hline
\end{tabular}
\caption{lOR of outcome by SNP class for first allele from formula \eqref{marginalmodel} (left number) and simulation (right number) varying prevalence of disease (first column). Case-control sampling design.}
\label{retroVaryPrevTable}
\end{center}
\end{table}
\begin{table}[ht]
\begin{center}
\begin{tabular}{c|cc|cc|cc|}
  \hline
 & all &   & novel &   & old &   \\ 
  \hline
0.25 & 0.03 & 0.02 & -0.03 & -0.02 & 0.03 & 0.03 \\ 
  0.5 & 0.11 & 0.10 & -0.11 & -0.09 & 0.14 & 0.12 \\ 
  0.75 & 0.24 & 0.21 & -0.23 & -0.18 & 0.29 & 0.26 \\ 
  1 & 0.40 & 0.35 & -0.36 & -0.28 & 0.48 & 0.43 \\ 
  1.25 & 0.58 & 0.52 & -0.49 & -0.37 & 0.68 & 0.61 \\ 
   \hline
\end{tabular}
\caption{lOR of outcome by SNP class for first allele from formula \eqref{marginalmodel} (left number) and simulation (right number) varying SD of SNP effects (first column). Case-control sampling design.}
\label{retroVaryTauTable}
\end{center}
\end{table}
\begin{table}[ht]
\begin{center}
\begin{tabular}{c|cc|cc|cc|}
  \hline
 & all &   & novel &   & old &   \\ 
  \hline
-0.47 & -0.25 & -0.28 & -0.56 & -0.52 & -0.20 & -0.23 \\ 
  0 & 0.21 & 0.18 & -0.20 & -0.16 & 0.26 & 0.23 \\ 
  0.26 & 0.46 & 0.44 & -0.01 & 0.04 & 0.51 & 0.48 \\ 
  0.47 & 0.66 & 0.64 & 0.13 & 0.20 & 0.70 & 0.67 \\ 
  0.69 & 0.87 & 0.85 & 0.28 & 0.39 & 0.91 & 0.88 \\ 
  0.92 & 1.09 & 1.07 & 0.43 & 0.58 & 1.11 & 1.09 \\ 
   \hline
\end{tabular}
\caption{lOR of outcome by SNP class for first allele from formula \eqref{marginalmodel} (left number) and simulation (right number) varying mean of SNP effects (first column). Case-control sampling design.}
\label{retroVaryMuTable}
\end{center}
\end{table}
\begin{table}[ht]
\begin{center}
\begin{tabular}{c|cc|cc|cc|}
  \hline
 & all &   & novel &   & old &   \\ 
  \hline
0.005 & 0.21 & 0.18 & -0.03 & -0.02 & 0.32 & 0.28 \\ 
  0.01 & 0.21 & 0.18 & -0.20 & -0.16 & 0.26 & 0.23 \\ 
  0.02 & 0.21 & 0.18 & -0.44 & -0.32 & 0.22 & 0.19 \\ 
  0.03 & 0.21 & 0.18 & -0.62 & -0.49 & 0.21 & 0.18 \\ 
  0.05 & 0.21 & 0.18 & -0.88 & -0.75 & 0.20 & 0.18 \\ 
   \hline
\end{tabular}
\caption{lOR of outcome by SNP class for first allele from formula \eqref{marginalmodel} (left number) and simulation (right number) varying SNP MAF (first column). Case-control sampling design.}
\label{retroVaryMafTable}
\end{center}
\end{table}
\begin{table}[ht]
\begin{center}
\begin{tabular}{c|cc|cc|cc|}
  \hline
 & all &   & novel &   & old &   \\ 
  \hline
0.005 & 0.21 & 0.19 & -0.03 & -0.02 & 0.32 & 0.29 \\ 
  0.01 & 0.21 & 0.18 & -0.20 & -0.16 & 0.26 & 0.23 \\ 
  0.02 & 0.21 & 0.16 & -0.44 & -0.31 & 0.22 & 0.17 \\ 
  0.03 & 0.21 & 0.15 & -0.62 & -0.46 & 0.21 & 0.15 \\ 
  0.05 & 0.21 & 0.12 & -0.88 & -0.65 & 0.20 & 0.12 \\ 
   \hline
\end{tabular}
\caption{lOR of outcome by SNP class for first allele from formula \eqref{marginalmodel} (left number) and simulation (right number) varying MAF and number of SNPs with constant expected count per person (first column). Case-control sampling design.}
\label{retroVaryMafNJointTable}
\end{center}
\end{table}
\begin{table}[ht]
\begin{center}
\begin{tabular}{c|cc|cc|cc|}
  \hline
 & all &   & novel &   & old &   \\ 
  \hline
50 & 0.21 & 0.20 & 0.07 & 0.08 & 0.37 & 0.32 \\ 
  100 & 0.21 & 0.20 & -0.03 & -0.02 & 0.32 & 0.29 \\ 
  200 & 0.21 & 0.20 & -0.20 & -0.17 & 0.26 & 0.25 \\ 
  500 & 0.21 & 0.20 & -0.54 & -0.43 & 0.21 & 0.20 \\ 
  1000 & 0.21 & 0.20 & -0.87 & -0.74 & 0.21 & 0.20 \\ 
   \hline
\end{tabular}
\caption{lOR of outcome by SNP class for first allele from formula \eqref{marginalmodel} (left number) and simulation (right number) varying sample size (first column). Case-control sampling design.}
\label{retroVarySampSizeTable}
\end{center}
\end{table}
\begin{table}[ht]
\begin{center}
\begin{tabular}{c|cc|cc|cc|}
  \hline
 & all &   & novel &   & old &   \\ 
  \hline
0.005 & 0.21 & 0.18 & -0.03 & -0.02 & 0.32 & 0.28 \\ 
  0.01 & 0.21 & 0.18 & -0.03 & -0.02 & 0.32 & 0.27 \\ 
  0.02 & 0.21 & 0.18 & -0.03 & -0.02 & 0.31 & 0.27 \\ 
  0.03 & 0.21 & 0.18 & -0.03 & -0.03 & 0.31 & 0.26 \\ 
  0.05 & 0.21 & 0.18 & -0.03 & -0.02 & 0.31 & 0.25 \\ 
   \hline
\end{tabular}
\caption{lOR of outcome by SNP class for first allele from formula \eqref{marginalmodel} (left number) and simulation (right number) varying MAF and sample size with constant expected count per SNP (first column = MAF). Case-control sampling design.}
\label{retroVarySampSizeImportantTable}
\end{center}
\end{table}


\end{document}